\documentclass[prd,11pt,floatfix,superscriptaddress,showkeys]{revtex4}

\usepackage{latexsym,amsbsy}
 \usepackage[dvips]{graphicx}
 \DeclareGraphicsExtensions{.eps, .jpg}
\begin{document}

\setcounter{page}{1}                                
\thispagestyle{empty}                                


\title{\boldmath
Cosmological models, observational data\\ and tension in Hubble constant }



\author{G. S. Sharov} 
\author{and E. S. Sinyakov}
\affiliation{Department of Mathematics, Tver State University,
Sadovyi per. 35, Tver, Russia}

\email{Sharov.GS@tversu.ru} 

\begin{abstract}
We analyze how predictions of cosmological models depend on a choice of
described observational data, restrictions on flatness,
 and how this choice can alleviate the $H_0$ tension. These
effects are demonstrated in the $w$CDM model in comparison with  the standard
$\Lambda$CDM model. We describe the Pantheon sample observations of Type Ia supernovae,
31 Hubble  parameter data points $H(z)$ from cosmic chronometers, the extended sample
with 57 $H(z)$ data points and observational manifestations of cosmic microwave
background radiation (CMB).
 For the $w$CDM and $\Lambda$CDM models in the flat case and with spatial curvature
we calculate $\chi^2$ functions for all observed data in different combinations,
estimate optimal values of model parameters and their expected intervals. For both
considered models the results essentially depend on a choice of data sets. In
particular, for the $w$CDM model with $H(z)$ data, supernovae and CMB the $1\sigma$
estimations may vary from $H_0=67.52^{+0.96}_{-0.95}$ km\,/(s$\cdot$Mpc) (for all
$N_H=57$ Hubble  parameter data points) up to   $H_0=70.87^{+1.63}_{-1.62}$
km\,/(s$\cdot$Mpc) for the flat case ($k=0$) and $N_H=31$. These results might be a hint
how to alleviate the problem of $H_0$ tension: different estimates of the Hubble
constant may be connected with filters and a choice of observational data.
\end{abstract}


 \maketitle \flushbottom


%
\section{Introduction}\label{Intr}

One of the most significant problem in modern cosmology is the tension between
estimations of the Hubble constant $H_0$ made (from one side) by Planck collaboration
during the last 6 years \cite{Planck13,Planck15,Planck18} with the recent fitting
 \cite{Planck18} $H_0=67.37\pm0.54$ km\,/(s$\cdot$Mpc) and  (from another side)
 by the Hubble Space Telescope (HST) group \cite{HST18,HST19}
 $H_0=74.03 \pm1.42$ km\,/(s$\cdot$Mpc).  Estimations of Planck collaboration are
based upon analysis of cosmic microwave background (CMB) data whereas the HST method
uses direct local distance ladder measurements
 of Cepheids in our Galaxy and in nearest galaxies, in particular,
observations of 70 Cepheids in the Large Magellanic Cloud in the latest paper
\cite{HST19}.

This mismatch between $H_0$ estimations of Planck and HST collaborations was not
diminishing but was growing during last years and now it exceeds $4\sigma$
\cite{Planck18,HST19}.

Cosmologists suggested different approaches for solving this problem: equations of state
with several variations, new components of matter, in particular, extra relativistic
species, modifications and transitions in early evolution, modifications of general
relativity, interactions of components and others
\cite{HuangWang2016}\,--\,\cite{DiValentMMVw2019} (see the extended list of literature
in Ref.~\cite{DiValentMMVw2019}). In particular, in papers
\cite{DiValentMM2017}\,--\,\cite{DiValentMMVw2019} scenarios with interaction between
dark energy and dark matter are explored. The authors analyze observational data with
these models and estimate optimal values of $H_0$, which can appear essentially
different (compatible with the tension described above) if they include or exclude the
interaction. The predicted value of $H_0$ in these scenarios is also sensitive to some
additional factors: curvature, neutrino masses, effective number of neutrino species,
variations in equation of state, etc.

In the present paper we demonstrate that similar variations of predicted values $H_0$
and their dependence on model parameters may be obtained in the (more simple) $w$CDM
model without interaction \cite{MotaB2003,BrevikNOV2004,NojiriOdin2005}. In this model
the dark energy component is described as a fluid with the equation of state
$p_x=w\rho_x$, $w={}$const. Other matter components (including the usual visible matter
and cold dark matter) in the $w$CDM scenario are the same as in the standard
$\Lambda$CDM model (see Sect.~\ref{Models}).

For the considered cosmological models estimations of the Hubble constant $H_0$ and
other model parameters are made via confronting the models with observational data. The
similar approach we used previously in papers \cite{GrSh13}\,--\,\cite{OdintsovSGS2019}.

In this paper we include in our analysis the following observations: the latest Type Ia
supernovae data (SNe Ia) from the Pantheon sample survey \cite{Scolnic17}, data
connected with cosmic microwave background radiation (CMB) and extracted from Planck
observations \cite{Planck15,HuangWW2015} and the Hubble parameter estimations $H(z)$ for
different redshifts $z$.

We analyze separately 31 Hubble  parameter data points $H(z)$ measured from differential
ages of galaxies (in other words, from cosmic chronometers), and the full set with 26
additional $H(z)$ data points obtained as observable effect of baryon acoustic
oscillations (BAO).  These data sets and effects of their choice were studied previously
in Ref.~\cite{SharovVas2018} for the model with generalized Chaplygin gas and the
$\Lambda$CDM model. All these 57 $H(z)$ data points were used in
Ref.~\cite{SharovV2018}, whereas in Ref.~\cite{OdintsovSGS2019} 31  $H(z)$ data points
from cosmic chronometers were applied to the $F(R)$ model considered there.

This paper is organized as follows. Details of dynamics and free model parameters for
the $w$CDM and $\Lambda$CDM scenarios are described in the next section.
Sect.~\ref{Observ} is devoted to $H(z)$,  SNe Ia and CMB  observational data, in
Sect.~\ref{Results} we analyze the results of our calculations for the $H(z)$ and  SNe
Ia observations, estimated values of model parameters  including  the Hubble constant
$H_0$ and in Sect.~\ref{SectCMB} we add to our analysis the CMB data.

\section{$\Lambda$CDM and $w$CDM models}
 \label{Models}

In the $\Lambda$CDM and $w$CDM models for a homogeneous isotropic Universe with the
Friedmann-Lema\^itre-Robertson-Walker line element
 \begin{equation}
ds^2 = -dt^2 + a^2 (t) \left[\frac{dr^2}{1-k r^2} + r^2 \left(d \theta^2 + \sin^2 \theta\,
d \phi^2\right)  \right]
  \label{FLRW} \end{equation}
 the Einstein equations are reduced to the system of 
the Friedmann equation
\begin{equation}
    3\frac{\dot{a}^2+k}{a^2}=8\pi G\rho+\Lambda
\label{Fried}
\end{equation}
and the continuity equation
\begin{equation}
    \dot\rho+3\frac{\dot a}{a}(\rho+p)=0.
    \label{cont}
\end{equation}

Here $a=a(t)$ is the scale factor, $\dot a=\frac {da}{dt}$ is its  derivative with
respect to  time $t$, $G$ is the Newton gravitational constant, $k$ is the sign of
spatial curvature, $\rho$ is the energy density of matter, $\Lambda$ is the cosmological
constant describing dark energy in the $\Lambda$CDM model; we choose the units where the
speed of light $c=1$.

In the $\Lambda$CDM and $w$CDM models the matter with density  $\rho$ in
Eq.~(\ref{Fried}) includes the cold matter component with density
$\rho_m=\rho_b+\rho_{dm}$ (it unifies baryons and dark matter, behaves like dust and has
zero pressure $p_m=0$),
 and the fraction of relativistic matter (radiation and neutrinos) with $\rho_r$
and pressure $p_r=\rho_r/3$. We suppose that the mentioned components and dark energy do
not interact in the form \cite{SharovBPNCh2017,PanSharov2017}, in other words, they
independently satisfy the continuity equation (\ref{cont}). We integrate this equation
with $p_m=0$ and $p_r=\rho_r/3$ and obtain the relations for cold and relativistic
matter:
\begin{equation}
    \rho_m=\rho_m^0\Big(\frac{a}{a_0}\Big)^{-3},\qquad
     \rho_r=\rho_r^0\Big(\frac{a}{a_0}\Big)^{-4}.
    \label{rhomr}
\end{equation}
Here the index ``0'' corresponds to the present time $t_0$, in particular,
$\rho_m^0=\rho_m(t_0)$, $a_0=a(t_0)$.

In sections below for both considered models we compare model predictions with
observations of the Hubble parameter
\begin{equation}
H=\frac {\dot{a}}{a}=\frac{d}{dt}\ln a.
    \label{H}
\end{equation}
We use observational data from our previous papers \cite{SharovVas2018,SharovV2018}
with estimations of $H=H (z)$ corresponding to
definite values of redshift $z$
\begin{equation}
z =\frac {\Delta\lambda}{\lambda} =\frac{a_0}{a} - 1.
 \label{z}
\end{equation}
 Parameter $z$ is observed with high accuracy as the ratio of a wavelength shift to an emitted wavelength.
In the relation (\ref{z}) $z+1 =a_0/a$ the scale factor $a$ corresponds to the event
(emission) epoch.

We express the Hubble parameter (\ref{H}) $H=H (a)$ or, equivalently, $H=H(z)$ from the
Friedmann equation (\ref{Fried}). For the $\Lambda$CDM model with density (\ref{rhomr})
and the $\Lambda$ term (describing the dark energy) this expression for the ratio of $H$
to the Hubble constant $H_0=H(t_0)$ takes the form
\begin{eqnarray}
\frac{H^2}{H_0^2}&=&\Omega_m^0\Big(\frac{a}{a_0}\Big)^{-3}+\Omega_r^0\Big(\frac{a}{a_0}\Big)^{-4}+
\Omega_\Lambda+\Omega_k\Big(\frac{a}{a_0}\Big)^{-2}, \label{H2}\\
&=&\Omega_m^0(1+z)^{3}+\Omega_r^0(1+z)^{4}+ \Omega_\Lambda+\Omega_k(1+z)^{2}.
\label{H2z}
\end{eqnarray}
Here
 \begin{equation}
\Omega_m^0=\frac{8\pi G\rho_m^0}{3H_0^2},\qquad\Omega_r^0=\frac{8\pi
G\rho_r^0}{3H_0^2},\qquad \Omega_\Lambda=\frac{\Lambda}{3H_0^2},\qquad
    \Omega_k=-\frac{k}{a_0^2H_0^2}
  \label{Om}
\end{equation}
are correspondingly fractions of cold matter ($\Omega_m^0$), radiation ($\Omega_r^0$),
dark energy ($\Omega_{\Lambda}$) and space-time curvature ($\Omega_k$) in the current
density balance.

Under the condition $z=0$ or $a=a_0$ (corresponding to the present time $t=t_0$)
the equations (\ref{H2}) or (\ref{H2z}) are reduced to the equality
 \begin{equation}
\Omega_m^0 + \Omega_r^0 + \Omega_{\Lambda} + \Omega_k = 1.
 \label{sumOm}
\end{equation}
Hence, the summands $\Omega_i$ in this equality are not independent. So we can consider
(any) three of these $\Omega_i$ as free parameters of the model.

One should note, that a large number of free model parameters is a disadvantage of any
cosmological scenario \cite{SharovBPNCh2017}\,--\,\cite{OdintsovSGS2019}. In order to
reduce the number of free parameters, we fix the radiation-matter ratio as provided by
Planck \cite{Planck13} in accordance with the previous papers
\cite{OdintsovSGS2017,OdintsovSGS2019}:
 \begin{equation}
X_r=\frac{\rho_r^0}{\rho_m^0}=\frac{\Omega_r^0}{\Omega_m^0}=2.9656\cdot10^{-4}\ .
\label{Xrm}
\end{equation}
 In other words, we fix  the effective number $N_{\mbox{\scriptsize eff}}$ of relativistic species
in accordance with the standard cosmological model and Planck data
\cite{Planck13,Planck15}: $N_{\mbox{\scriptsize eff}}= 3.046\pm0.18 $. Because of small
value $X_r$ the relativistic (radiation) fraction $\Omega_r$ is insufficient for $H(z)$
and Type Ia Supernovae observational data concerning redshifts $0\le z\le2.36$. In
Sect.~\ref{Observ} we shall apply this component with its fraction
$\Omega_r(z)=\Omega_r^0(1+z)^{4}$ to describing observational manifestations of cosmic
microwave background radiation (CMB) with the fixed value $X_r$ (\ref{Xrm}).

Under the condition (\ref{Xrm}) the $\Lambda$CDM model (describing the late time
evolution of the Universe) has three independent parameters: $H_0$ and any two of the
three $\Omega_i$. Below we use $\Omega_m^0$ and $\Omega_k$ as independent parameters.

The $w$CDM model generalizes the $\Lambda$CDM scenario. In the $w$CDM model the cold and
relativistic matter components are just the same (with evolution (\ref{rhomr}) of
densities $\rho_m=\rho_b+\rho_{dm}$ and $\rho_r$), but the dark energy is described as a
fluid, whose pressure $p_x$ is related to the energy density $\rho_x$ by the ratio
$p_x=w\rho_x$. Here the constant $w$ is the additional free parameter in this model, where
$\Lambda=0$ and the total energy density  is $\rho=\rho_m+\rho_r+\rho_x$.

Thus, from Friedmann equation (\ref{Fried}) we deduce the analog of the Eq. (\ref{H2})
or (\ref{H2z}) for the $w$CDM model:
 \begin{equation}
 \frac{H^2}{H_0^2}=\Omega_m^0(1+z)^{3}+\Omega_r^0(1+z)^{4}+\Omega_k(1+z)^{2}+ \Omega_x(1+z)^{3(1+w)}.
  \label{H2w}
\end{equation}
Here the dark energy fraction $\Omega_x^0 =8\pi G\rho_x^0/(3H_0^2)$ is connected with
other fractions
$$\Omega_m^0 + \Omega_r^0 + \Omega_x^0 + \Omega_k = 1.$$
This analog of Eq.~(\ref{sumOm}) results from equation (\ref{H2w}) at $z=0$.

Hence, in the $w$CDM model we have four independent parameters,  we should add $w$ to
the set of three known ($\Lambda$CDM)  parameters: $H_0$, $\Omega_m^0$ and $\Omega_k$.

In the particular case $w = -1$ the $w$CDM model (\ref{H2w}) transforms into the
$\Lambda$CDM model (\ref{H2z}).

\section{Observational data}\label{Observ}


As was mentioned above, for the considered cosmological models we calculate their
optimal model parameters taking into account the best correspondence to a chosen set of
observational data. These data include: 1) estimates of the Hubble parameter $H(z)$ at
various redshifts; 2) observations of Type Ia supernovae (SNe Ia) from the Pantheon
sample \cite{Scolnic17} and 3) data from Planck observations of cosmic microwave
background radiation (CMB) \cite{Planck15,HuangWW2015}.

In accordance with the previous papers \cite{ShV14}\,--\,\cite{OdintsovSGS2019} we
divide the Hubble parameter data  $H(z)$ into two parts. The first part contains now 31
estimations of $H(z)$ (named also cosmic chronometers) measured  via differential ages
of galaxies $\Delta t$, the formula (\ref{z}) and its corollary
$$ 
 H (z)= \frac{\dot{a}}{a}= -\frac{1}{1+z}
\frac{dz}{dt} \simeq -\frac{1}{1+z}
\frac{\Delta z}{\Delta t}.
 $$

The second method uses observations based on baryon acoustic oscillation (BAO) data
along the line-of-sight directions. In this paper we use 31 $H(z)$ data points from
cosmic chronometers and 26 data points obtained with BAO method, all these data and
corresponding references are tabulated in Refs.~\cite{SharovVas2018,SharovV2018}.

We analyze separately $N_H=31$ $H(z)$ data points from cosmic chronometers, and the full
set with all $N_H=57=31+26$ Hubble parameter data points. For a cosmological model with
free parameters denoted by $p_1, p_2,\dots$, the best fitted (optimal) values of $p_j$
with respect to the $H(z)$  observational data are achieved, if the $\chi^2$ function
\cite{ShV14}\,--\,\cite{OdintsovSGS2019}
\begin{equation}
    \chi_H^2(p_1,p_2,\dots)=\sum_{j=1}^{N_H}\bigg[\frac{H(z_j,p_1,p_2,\dots)-H^{obs}(z_j)}{\sigma _j}  \bigg]^2,
    \label{chiH}
\end{equation}
 reaches its minimum in this parameter space. Here $N_H$ is the number of observations,
$H^{obs}(z_j)$ are observational data with errors $\sigma_j$, $H(z_j,p_1,p_2,\dots)$ are
theoretical values of Hubble parameter (\ref{H}) calculated from Eqs.~(\ref{H2z}) or
(\ref{H2w}) for the $\Lambda$CDM or $w$CDM model correspondingly.

In the next section we shall demonstrate that the analysis of only Hubble parameter data
and the function $\chi_H^2$ is not reliable enough for these cosmological models. We
should include into consideration the Type Ia supernovae data.

Observations of Type Ia supernovae were the first evidence of accelerated expansion of
the Universe, they play an essential role in striking progress of cosmology during the
last two decades \cite{NojOdinFR,BambaCNO12}. Supernovae are stars which explode with
release of huge energy and expanding their outer shell. These objects are classified in
correspondence with their spectrum and time evolution of their brightness
\cite{SNKirshner09}. The most interesting class of them is Type Ia supernovae, which are
usually considered as standard candles in the Universe, because we can determine their
epoch (redshift $z$) and the distance (luminosity distance $D_L$) to these objects. The
luminosity distance \cite{ShV14}\,--\,\cite{OdintsovSGS2019}
\begin{equation}
 D_L(z)=\frac{c\,(1+z)}{H_0}S_k
 \bigg(H_0\int\limits_0^z\frac{d\tilde z}{H(\tilde
 z)}\bigg),  \label{DL}
\end{equation}
depends on the sign $k$ of spatial curvature of the  FLRW Universe (\ref{FLRW}) via the
expression
$$
S_k(x)=\left\{\begin{array}{ll} \sinh\big(x\sqrt{\Omega_k}\big)\big/\sqrt{\Omega_k}, &\Omega_k>0,\\
 x, & \Omega_k=0,\\ \sin\big(x\sqrt{|\Omega_k|}\big)\big/\sqrt{|\Omega_k|}. &
 \Omega_k<0,
 \end{array}\right.
 $$
 Here $\Omega_k$ is the curvature fraction (\ref{Om}).

In papers \cite{GrSh13}\,--\,\cite{SharovV2018} we used the Union 2.1 table
\cite{SNTable}, containing 580 observations of Type Ia supernovae (SNe Ia), however in
Ref.~\cite{OdintsovSGS2019} and in this paper we use the Pantheon sample
\cite{Scolnic17}, that is the latest (2017) extended SNe Ia data set, containing
information about $N_{\mbox{\scriptsize SN}}=1048$ Type Ia supernovae. This information
includes the redshift values $z=z_i$ of objects, their luminosity distance moduli
(logarithms of the luminosity distance $D_L$)
$$\mu_i=\mu(D_L)=5\lg\big(D_L/10\mbox{pc}\big),$$
and the $N_{\mbox{\scriptsize SN}}\times N_{\mbox{\scriptsize SN}}$ covariance matrix
$C_{\mbox{\scriptsize SN}}$ for these data points.

The observed values $\mu_i=\mu_i^{obs}$ with the inverse  matrix $C_{\mbox{\scriptsize
SN}}^{-1}$ from the Pantheon sample \cite{Scolnic17} and the theoretically deduced
Hubble parameter $H(z)=H(z,p_1,\dots)$ (\ref{H2z}) or (\ref{H2w}) let us calculate the
functions $D_L(z)$ (\ref{DL}), $\mu(z)=\mu^{th}(z,p_1,\dots)$ and the corresponding
$\chi^2$ function for SNe Ia data \cite{OdintsovSGS2019}:
 \begin{equation}
 \chi^2_{\mbox{\scriptsize SN}}(p_1,\dots)=\min_{H_0}\sum_{i,j=1}^{N_{\mbox{\scriptsize SN}}}
 \Delta\mu_i\big(C_{\mbox{\scriptsize SN}}^{-1}\big)_{ij} \Delta\mu_j,\;
\Delta\mu_i=\mu^{th}(z_i,p_1,\dots)-\mu_i^{obs},
  \label{chiSN}
 \end{equation}
 Here $p_1, p_2,\dots=\Omega_m^0, \Omega_k,\dots$ are free parameters of the $\Lambda$CDM or $w$CDM
models. To eliminate data errors, in the formula (\ref{chiSN}) we should minimize
(marginalize) over $H_0$, so the resulting function $ \chi^2_{\mbox{\scriptsize
SN}}(\Omega_m^0,\dots)$ does not depend on $H_0$.

In the next section we study how the $\Lambda$CDM or $w$CDM models describe the unified
set of observational data, including observation of the Hubble parameter  $H(z)$ and
Type Ia supernovae. The results are determined by the $\chi^2$ function, that is the sum
of the functions (\ref{chiH}) and (\ref{chiSN}):
\begin{equation}
 \chi^2_{H+\mbox{\scriptsize SN}}(p_1,\dots)= \chi^2_{H}(p_1,\dots)+\chi^2_{\mbox{\scriptsize SN}}(p_1,\dots).
  \label{chiHSN}
\end{equation}

In this paper (unlike Refs.~\cite{ShV14}\,--\,\cite{OdintsovSGS2019}) we do not include
into consideration manifestations of baryon acoustic oscillations (BAO) to avoid
correlation with and 26 $H(z)$ data points obtained with BAO method.

However, in accordance with Refs.~\cite{OdintsovSGS2017,OdintsovSGS2019} we investigate
in Sect.~\ref{SectCMB} changes in model predictions from observational manifestations of
cosmic microwave background radiation (CMB). We use the CMB observational parameters
\cite{HuangWW2015}
  \begin{equation}
  \mathbf{x}=\big(R,\ell_A,\omega_b\big);\qquad R=\sqrt{\Omega_m^0}\frac{H_0D_M(z_*)}c,\quad
 \ell_A=\frac{\pi D_M(z_*)}{r_s(z_*)},\quad\omega_b=\Omega_b^0h^2,
 \label{CMB} \end{equation}
 related with the photon-decoupling epoch $z_*=1089.90 \pm0.25$ \cite{Planck13,Planck18}
 (unlike the SNe Ia and $H(z)$, measured for
$0<z\le2.36$). Here $D_M(z)=D_L(z)\big/(1+z)$,
 $h=H_0/[100\,\mbox{km}\mbox{s}^{-1}\mbox{Mpc}^{-1}]$,
 the comoving sound horizon $r_s$ at $z=z_*$ is calculated as
  $$
  r_s(z)=\frac1{\sqrt{3}}\int_0^{1/(1+z)}\frac{da}
 {a^2H(a)\sqrt{1+\big[3\Omega_b^0/(4\Omega_r^0)\big]a}}\ .
   $$
 In these calculations at high redshifts radiation is essential,
 so we use the fixed radiation-matter ratio $X_r=\Omega_r^0/\Omega_m^0$  in the form
(\ref{Xrm}). We consider the current baryon fraction $\Omega_b^0$  as the nuisance
parameter and marginalize over $\omega_b=\Omega_b^0h^2$ the  following $\chi^2_{\mbox{\scriptsize CMB}}$
function:
 \begin{equation}
\chi^2_{\mbox{\scriptsize CMB}}=\min_{\omega_b}\Delta\mathbf{x}\cdot
C_{\mbox{\scriptsize CMB}}^{-1}\big(\Delta\mathbf{x}\big)^{T},\qquad \Delta
\mathbf{x}=\mathbf{x}-\mathbf{x}^{Pl}\ .
 \label{chiCMB} \end{equation}
 We use the data \cite{HuangWW2015}
  \begin{equation}
  \mathbf{x}^{Pl}=\big(R^{Pl},\ell_A^{Pl},\omega_b^{Pl}\big)=\big(1.7448\pm0.0054,\;301.46\pm0.094,\;0.0224\pm0.00017\big)
   \label{CMBpriors} \end{equation}
extracted from Planck collaboration \cite{Planck15} with free amplitude for the lensing
power spectrum. The covariance matrix $C_{\mbox{\scriptsize CMB}}=\|\tilde
C_{ij}\sigma_i\sigma_j\|$ and other details are described in
Refs.~\cite{OdintsovSGS2017,OdintsovSGS2019} and \cite{HuangWW2015}.

\section{Analysis of $H(z)$ and  SNe Ia data}\label{Results}

We  begin our investigation from the analysis of the Hubble parameter data $H(z)$ and
the corresponding function  (\ref{chiH}) $\chi^2_H(\Omega_m^0,\dots)$, depending on
$\Omega_m^0,\Omega_k,H_0$ for the $\Lambda$CDM model and  on $\Omega_m^0,\Omega_k,H_0,w$
for the $w$CDM model.

\begin{figure}[th]
  \centerline{\includegraphics[scale=0.71,trim=4mm 4mm 5mm 5mm]{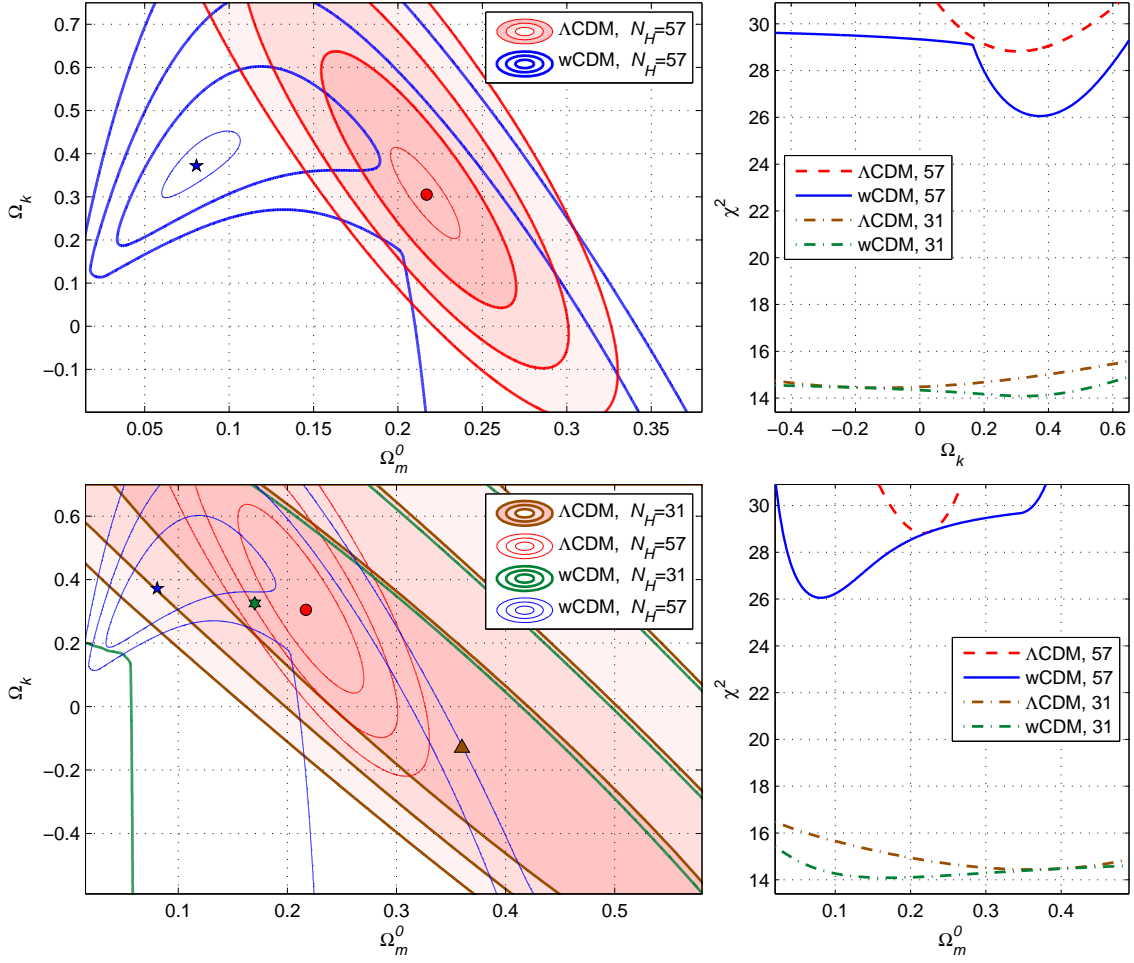}}
\caption{Contour plots of $\chi^2_H(\Omega_m^0,\Omega_k)$ with $N_H=57$ (the top-left
panel) and  with $N_H=31$ (the bottom-left panel) at $1\sigma$, $2\sigma$ and $3\sigma$
CL for the $\Lambda$CDM (filled contours) and $w$CDM models, the correspondent
one-parameter distributions are presented in the right panels. }
  \label{F1}
\end{figure}

In Fig. \ref{F1} we compare contour plots of  $\chi^2_H$ for these two models for all
$N_H=57$ Hubble parameter data points and for $N_H=31$ data points from cosmic
chronometers in the $\Omega_m^0-\Omega_k$ plane, more precisely, we draw the contour
plots at $1\sigma$ (68.27\%), $2\sigma$ (95.45\%) and $3\sigma$ (99.73\%) confidence
level for the two-parameter distributions
\begin{equation}
 \chi^2_H(\Omega_m^0, \Omega_k)=\left\{\begin{array}{ll}
 \min\limits_{H_0} \chi^2_H(\Omega_m^0, \Omega_k, H_0), & \mbox{ for $\Lambda$CDM},\\
 \min\limits_{H_0,w} \chi^2_H(\Omega_m^0,\Omega_k, H_0,w), & \mbox{ for $w$CDM}.
 \end{array}\right.
 \label{chiHLw1}
\end{equation}

In the top-left panel of Fig. \ref{F1} we show and compare these contour plots (thick
lines)  for both models for the case $N_H=57$. In the bottom-left panel we consider the
case $N_H=31$ (thick lines) and compare them from the previous contours for $N_H=57$
(thin lines with the same colors).

The corresponding one-parameter distributions $\chi^2_H (\Omega_m^0)$ and $\chi^2_H
(\Omega_k)$ (where $\chi^2_H$ is minimized over all other parameters) are shown at the
right panels of Fig.~\ref{F1}.

In the contour plots in Fig.~\ref{F1} positions of $\chi^2_H$ minima are shown as the
red circle and brown triangle for the $\Lambda$CDM model with, correspondingly, $N_H=57$
and 31;  and as the blue pentagram or green hexagram for the $w$CDM model. These colors
and marks will also be used below.  One can see the large difference between positions
of these minima points, especially for $\Omega_m^0$ with $N_H=57$ (observed  in the
top-left and bottom-right panels), the optimal values are: $\Omega_m^0\simeq0.217$ for
the $\Lambda$CDM and $\Omega_m^0\simeq0.081$ for the $w$CDM model. The last value
strongly differs from modern estimates of this parameter $\Omega_m^0\simeq0.3$
\cite{Planck15,Planck18}.

In addition, if we use only the Hubble parameter $H(z)$ data, the optimal values of the
curvature fraction $\Omega_k$ in 3 considered cases of models and $N_H$ are larger than
$0.3$, but this value is negative for the $\Lambda$CDM with $N_H=31$. The positive
($\Omega_k>0$) $1\sigma$ domains for both models in the case $N_H=57$ essentially exceed
the close to zero limits $\Omega_k=0.0007\pm0.0037$, coming from the latest multivariate
estimations \cite{Planck18}. For the case $N_H=31$ both model predict the best fitted
values $\Omega_k$ with different signs (strongly separated), however these estimates do
not exclude $\Omega_k\simeq 0$ values because of large  $1\sigma$ errors (see
Table~\ref{Estim}): $\Omega_k=-0.13^{+0.72}_{-0.54}$ for the  $\Lambda$CDM and
$\Omega_k=0.325^{+0.367}_{-1.96}$ for the $w$CDM model with $N_H=31$.

On can see also the non-standard behavior of the  contour plots (and the graph $\chi^2_H
(\Omega_k)$ in the bottom-left panel) in Fig.~\ref{F1} for the $w$CDM model, these lines
are bent. This effect appears, because at some points of the $ \Omega_m^0-\Omega_k$
plane, when we fix $\Omega_m^0$ and $\Omega_k$, the function $\chi^2_H$  of two
remaining parameters $H_0$, $w$ can have two local minima, and we should chouse the
minimal one from them (coinciding with the global minimum). This ``competition'' between
local minima is seen in Fig.~\ref{F1} at points, where we ``switch'' from one local
minimum to another during the minimization procedure in the expression (\ref{chiHLw1})
for $\chi^2_H(\Omega_m^0, \Omega_k)$. This effect should be carefully taken into
account. Note that the $\Lambda$CDM model has no such a  behavior (see Fig.~\ref{F1}).

\begin{figure}[th]
  \centerline{\includegraphics[scale=0.68,trim=6mm 4mm 5mm 1mm]{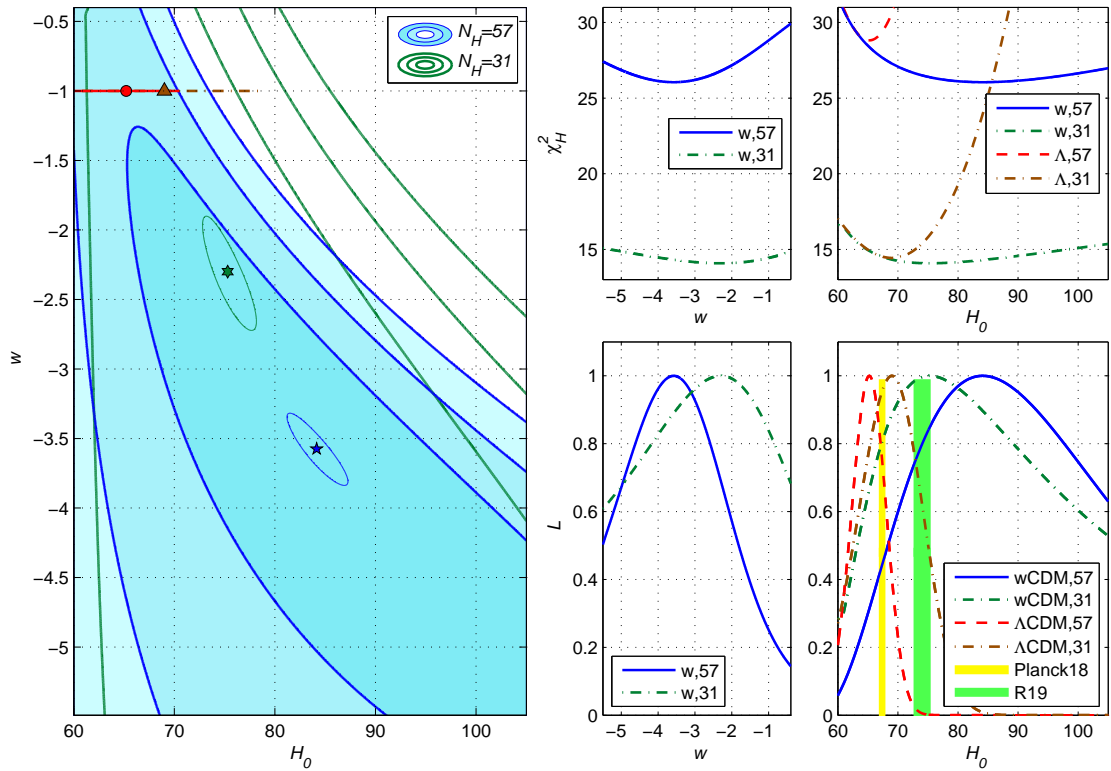}}
\caption{Contour plots of $\chi^2_H(H_0,w)$ for the $w$CDM model with $N_H=57$ (filled
contours) and with  $N_H=31$ (green contours), one-parameter distributions and
likelihood function ${\cal L}_H(w)$ and ${\cal L}_H(H_0)$ are shown in the right panels.
In the bottom-right panel the vertical bands refer to $H_0$ estimates of Planck 2018
\cite{Planck18} (yellow) and HST \cite{HST19} (green, labeled as R19).}
  \label{F2}
\end{figure}

For the $w$CDM model in Fig.~\ref{F2} we consider (filled) contour plots for the
two-parameter distribution in the $H_0-w$ plane:
$\chi^2_H(H_0,w)=\min\limits_{\Omega_m^0,\Omega_k} \chi^2_H$. Here we  use the same
notation. However, predictions of the $\Lambda$CDM model in this plane are contracted to
the $w=-1$ level line. In the right panels we show the  one-parameter distributions
$\chi^2_H(w)$ and $\chi^2_H (H_0)$ (minimized over all other parameters) and the
correspondent likelihood functions, in particular,
 \begin{equation}
{\cal L}_H(H_0)\sim\exp(-\chi^2_H(H_0)/2)
  \label{likeli}
\end{equation}
 We use these  functions for estimating $1\sigma$ errors, they are tabulated below
in Table~\ref{Estim}  with the best fitted values of the model parameters and minimums
of $\chi^2_H$ (they are $28.82$ for  the $\Lambda$CDM and $26.05$ for the $w$CDM model
for $N_H=57$).

In the bottom-right panel of Fig.~\ref{F2} we draw the vertical bands describing
correspondingly the $H_0$ estimates of Planck 2018 \cite{Planck18} and HST \cite{HST19}
(labeled here and below as Planck18 and R19). These bands and  $1\sigma$ estimates in
different models are reproduced below in Fig.~\ref{FWh} in the whisker plots. One can see
that for the case with $N_H=31$ Hubble parameter data points the $\Lambda$CDM $\chi^2_H$
prediction (the best fitted value) $H_0=69.0^{+5.15}_{-5.5}$ is close to Planck18 and
the $w$CDM estimation $75.3^{+24.5}_{-10.8}$ corresponds to R19, so it seems (at the
first glance) that we solve the $H_0$ tension problem, if we just switch from the
$\Lambda$CDM to $w$CDM predictions under these assumptions (only $\chi^2_H$ with 31
$H(z)$ data points).

However, other optimal values of model parameters under the mentioned assumptions (see
Table~\ref{Estim}), in particular,  the $w$CDM ($N_H=57$) estimations
$\Omega_k=0.372^{+0.149}_{-0.13}$ are far beyond the observational limits
\cite{Planck15,Planck18}.

{\small

\begin{table}[ht]
\caption{Optimal values and $1\sigma$ estimates of model parameters for $H(z)$ data}
\begin{center}
\begin{tabular}{||c|c|c|c|c|c|c|c||}
\hline Model & Data&$N_H$ &$\min\chi^2$ \rule{0pt}{1.4em}& $H_0$ & $\Omega_m^0$  & $\Omega_k$ & $w$  \\
\hline
$\Lambda$CDM&$H(z)$&31 & 14.44 & $69.0^{+5.15}_{-5.5}$ & $0.360^{+0.204}_{-0.233}$& $-0.13^{+0.72}_{-0.54}$ &  $-1$ \rule{0pt}{1.4em}  \\
\hline
$w$CDM  & $H(z)$  & 31 & 14.09 & $75.3^{+24.5}_{-10.8}$& $0.170^{+0.425}_{-0.134}$& $0.325^{+0.367}_{-1.96}$ & $-2.30^{+2.20}_{-3.22}$\rule{0pt}{1.4em}  \\
\hline
$\Lambda$CDM&$H(z)$&57 & 28.82 & $65.25^{+2.8}_{-2.9}$ & $0.217^{+0.036}_{-0.040}$& $0.305^{+0.209}_{-0.18}$ &  $-1$\rule{0pt}{1.4em}  \\
\hline
$w$CDM  & $H(z)$ &  57 & 26.05 & $84.2^{+21.8}_{-14.05}$&$0.081^{+0.054}_{-0.035}$& $0.372^{+0.149}_{-0.13}$ & $-3.57^{+1.49}_{-1.62}$\rule{0pt}{1.4em}  \\
\hline
 \end{tabular}
\end{center}
 \label{Estim}
\end{table}
}

Moreover, the  best fitted $\chi^2_H$ values in the case $N_H=57$:
$H_0=65.25^{+2.8}_{-2.9}$ from the $\Lambda$CDM and  $84.2^{+21.8}_{-14.05}$ from the
$w$CDM estimations have more larger spread than the tension between Planck18 and R19.
These estimations are also illustrated in the whisker diagram in Fig.~\ref{FWh},
corresponding the bottom-right panel of Fig.~\ref{F2} in comparison with the results,
determined by the function $\chi^2_{H+\mbox{\scriptsize SN}}$.

\begin{figure}[bh]
  \centerline{\includegraphics[scale=0.7,trim=2mm 4mm 5mm 1mm]{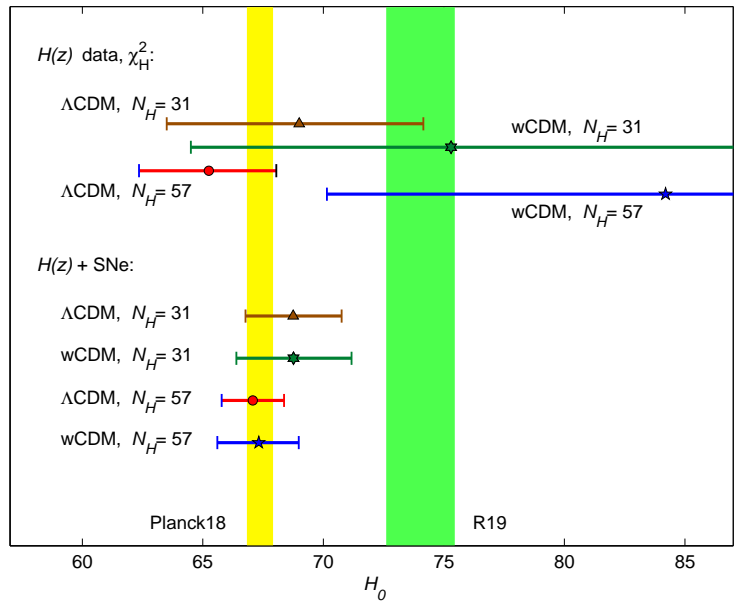}}
\caption{  Whisker plots for $\chi^2_H$,  $\chi^2_{H+\mbox{\scriptsize SN}}$ and 2
models with different $N_H$ in comparison with  Planck18 and R19 $H_0$ estimates. }
  \label{FWh}
\end{figure}


\begin{figure}[th]
  \centerline{\includegraphics[scale=0.7,trim=2mm 4mm 5mm 6mm]{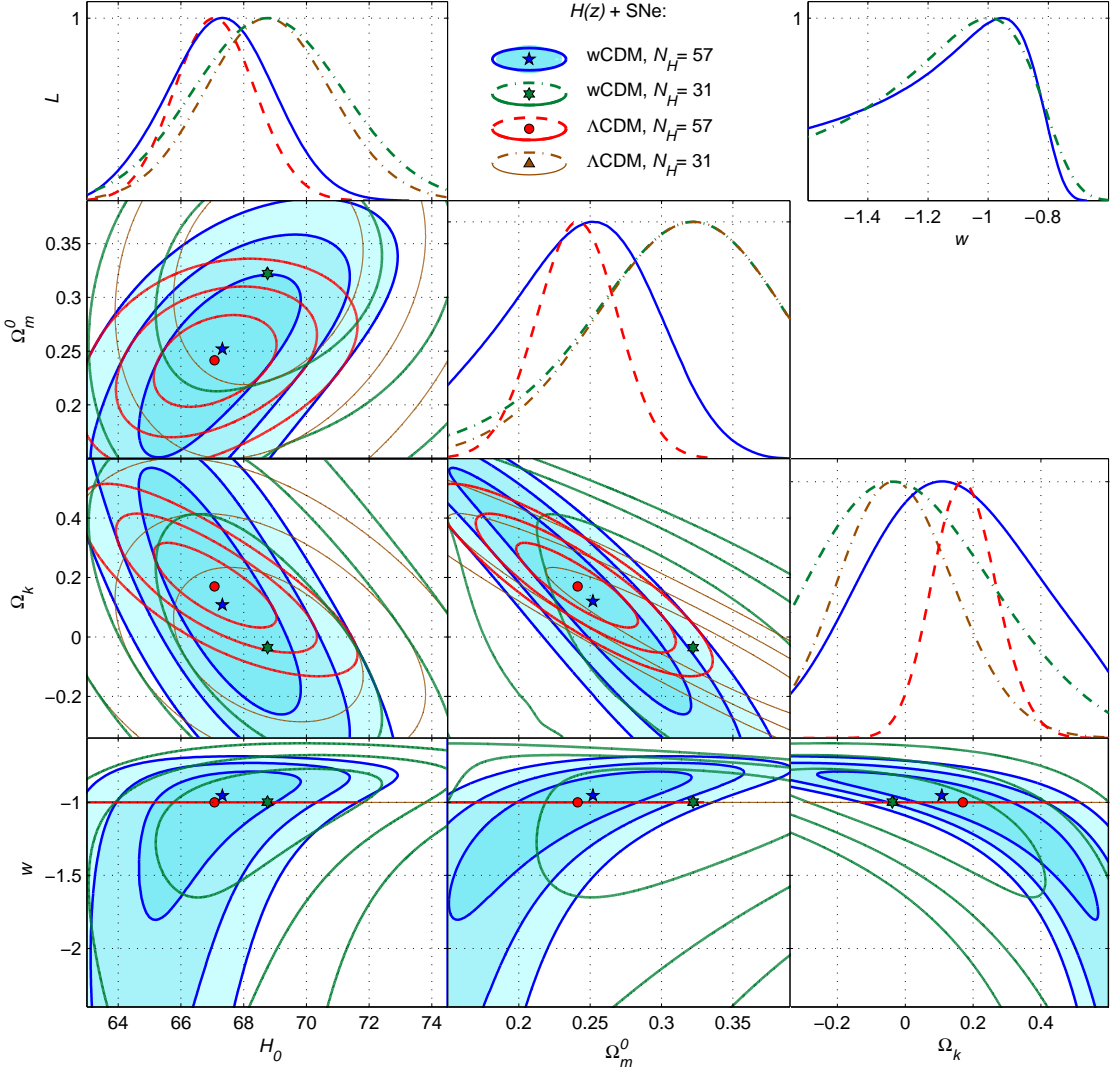}}
\caption{  Contour plots and one-parameter distributions of $\chi^2_{H+\mbox{\scriptsize
SN}}$ ($H(z)$ and the Pantheon  SNe Ia data) for the  $\Lambda$CDM and $w$CDM models. }
  \label{F4}
\end{figure}

Keeping in mind the non-standard estimations of $H_0$, $\Omega_k$ and behavior in the
$\Omega_m^0-\Omega_k$ and $H_0-w$ planes, one may conclude, that the Hubble parameter
observations $H(z)$ alone do not give an adequate picture of  the $\Lambda$CDM and
$w$CDM cosmology during $0\le z\le2.36$. Hence, we should add other observational data
described above in Sect.~\ref{Observ}, in particular, SNe Ia data \cite{Scolnic17}.

We consider further the  $H(z)$ with SNe Ia  data set described by the function
$\chi^2_{H+\mbox{\scriptsize SN}}= \chi^2_{H}+\chi^2_{\mbox{\scriptsize SN}}$
({\ref{chiHSN}}): the results are depicted in  Fig.~\ref{F4}, where we compare the
$\Lambda$CDM and $w$CDM models in 6 planes with contour plots ($H_0-\Omega_m^0$,
$H_0-\Omega_k$, $\Omega_m^0-\Omega_k$, $\Omega_m^0-\Omega_k$, etc.) and in 4 panels with
one-parameter  likelihood functions ${\cal L}_{H+\mbox{\scriptsize SN}}(p_j)$ of the
type (\ref{likeli}).  In all panels the blue filled contours and blue lines correspond
to $\chi^2_{H+\mbox{\scriptsize SN}}$ for the $w$CDM model with $N_H=57$, colors and
labels of minima points for other variants are the same as in Figs.~\ref{F1} and
\ref{F2}.

The one-parameter likelihood functions ${\cal L} _{H+\mbox{\scriptsize SN}}$ in
Fig.~\ref{F4} let us calculate the best fitted values and corresponding error bands
presented in Table~\ref{Estim2}.

One may observe in Fig.~\ref{F4} that the Pantheon  SNe Ia data, included in our
analysis, significantly change the best fitted values for all parameters and all
variants of the models (supporting the above mentioned irrelevance of only Hubble
parameter data). These values for $\chi^2_{H+\mbox{\scriptsize SN}}$ are tabulated in
Table~\ref{Estim2}. In particular, the exotic  $\chi^2_H$ estimates for the $w$CDM
($N_H=57$) model  $\Omega_m^0=0.081^{+0.054}_{-0.035}$, $w=-3.57^{+1.49}_{-1.62}$ in the
case $\chi^2_{H+\mbox{\scriptsize SN}}$ return to their ``normal'' values (corresponding
to recent  estimates \cite{Planck15,Planck15}): $\Omega_m^0=0.252^{+0.048}_{-0.061}$,
$w=-0.954^{+0.124}_{-0.33}$. The similar changes (up to
$\Omega_m^0=0.322^{+0.066}_{-0.069}$) take place also for the $w$CDM model with
$N_H=31$; in this case the optimal $w$CDM value $w=-0.988^{+0.166}_{-0.32}$ appears to
be extremely close the  $\Lambda$CDM limit $w=-1$ and the best fitted estimates of all
parameters practically coincide for these two models.

Fig.~\ref{F4} also demonstrates  the large difference between the $\Omega_m^0$ estimates
for the cases $N_H=31$ and 57. The similar  difference may be seen for the Hubble
parameter  $H_0$, however the whisker plot in Fig.~\ref{FWh} shows, that it is
essentially less than for the only  $\chi^2_H$ data. Thus, one may conclude, that for
the Hubble parameter plus SNe Ia data ($\chi^2_{H+\mbox{\scriptsize SN}}$) in all 4
considered variants of the  $\Lambda$CDM and $w$CDM models only  Planck18 estimates of
$H_0$ are supported: all models are in tension with the HST (R19) data.

\section{Additional analysis  of CMB data}
\label{SectCMB}

\begin{figure}[bh]
  \centerline{\includegraphics[scale=0.68,trim=2mm 4mm 5mm 0mm]{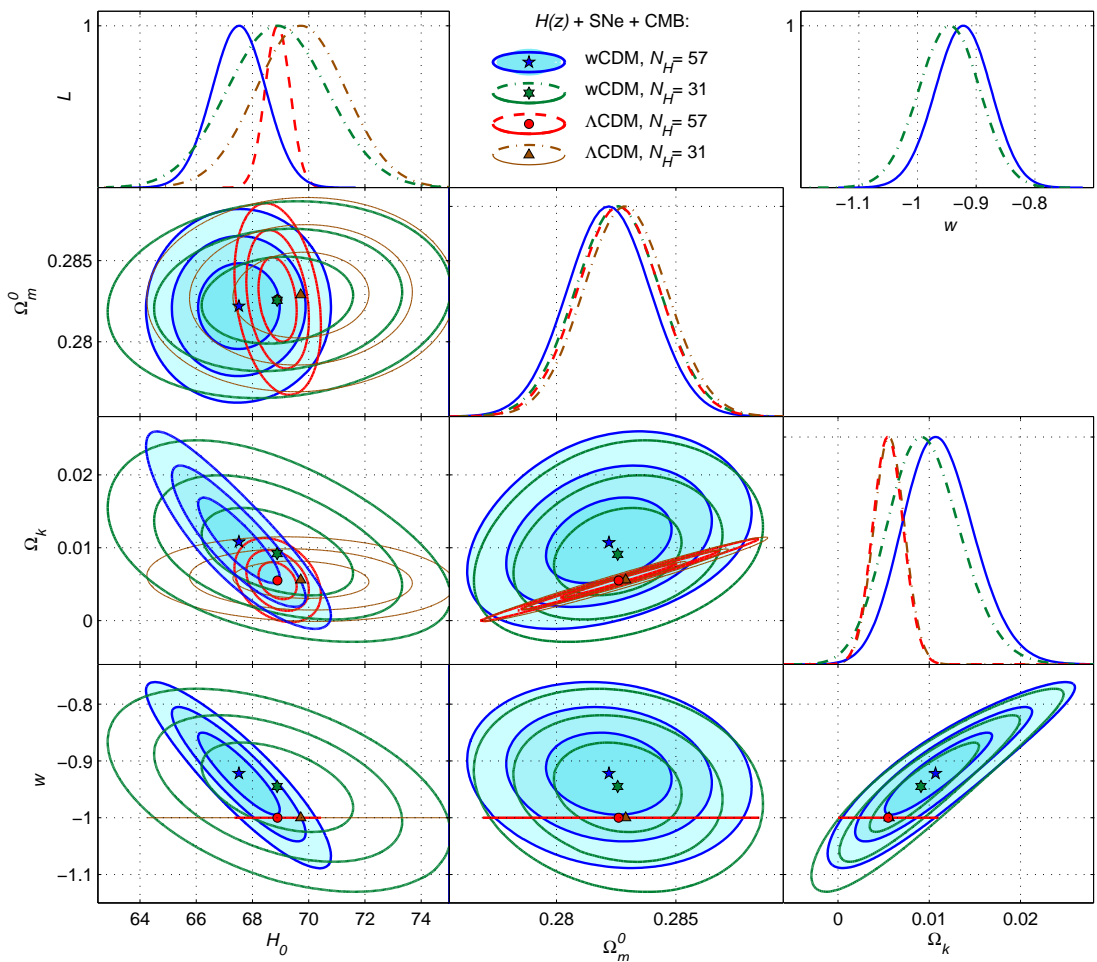}}
\caption{  Contour plots and one-parameter distributions of $\chi^2_{\mbox{\scriptsize
tot}}$ ($H+{}$SNe Ia${}+{}$CMB data) for the  $\Lambda$CDM and $w$CDM models. }
  \label{F5}
\end{figure}

In this section we add the cosmic microwave background radiation (CMB) data in the form
$\chi^2_{\mbox{\scriptsize CMB}}$ (\ref{chiCMB}), (\ref{CMBpriors})  \cite{HuangWW2015}
to the previous $H(z)$ and SNe Ia data sets and analyze the resulting $\chi^2$ function
 \begin{equation}
\chi^2_{\mbox{\scriptsize tot}}=\chi^2_H+\chi^2_{\mbox{\scriptsize
SN}}+\chi^2_{\mbox{\scriptsize CMB}}.
 \label{chitot} \end{equation}
 The results of $\chi^2_{\mbox{\scriptsize tot}}$-based calculations are presented in
Fig.~\ref{F5} and in Table~\ref{Estim2}.

One can expect from the previous papers \cite{OdintsovSGS2017,OdintsovSGS2019} (and will
see in Fig.~\ref{F5}) that the included CMB data strongly change estimations for model
parameters and especially for their error boxes. In particular, calculated from
$\chi^2_{\mbox{\scriptsize tot}}$ error boxes for $\Omega_m^0$ are essentially more
narrow because of small errors $\sigma_i$ in the CMB priors (\ref{CMBpriors}) of the
values (\ref{CMB}) with the parameter $R$ proportional to $\sqrt{\Omega_m^0}$.

In  Fig.~\ref{F5} and in Table~\ref{Estim2} we can observe, that the predicted from
$\chi^2_{\mbox{\scriptsize tot}}$ ($H+{}$SNe Ia${}+{}$CMB) error bands are strongly
contracted (in comparison with $\chi^2_{H+\mbox{\scriptsize SN}}$) not only for
$\Omega_m^0$ (where the error box is of order $\Delta\Omega_m^0\simeq0.0017$),
 but also for $\Omega_k$, where $\Delta\Omega_k\simeq0.0017$ for the $\Lambda$CDM and
$\Delta\Omega_k\simeq0.004$ for the $w$CDM model. One should note also, that the best
fitted estimates of $\Omega_m^0$ with the CMB data are rather close in the range
$0.282<\Omega_m^0<0.283$ for all 4 considered variants. For $\Omega_k$ the optimal
values lie in the range $0.0055\le\Omega_k\le0.011$ and slightly differ for the
$\Lambda$CDM and $w$CDM models.

However, for the Hubble constant the influence of  the CMB data is not so striking: the
$H_0$ error bands for $\chi^2_{\mbox{\scriptsize tot}}$  appear to be about $1.5$ times
diminished in comparison with the case.

\begin{figure}[bh]
  \centerline{\includegraphics[scale=0.7,trim=2mm 4mm 5mm 1mm]{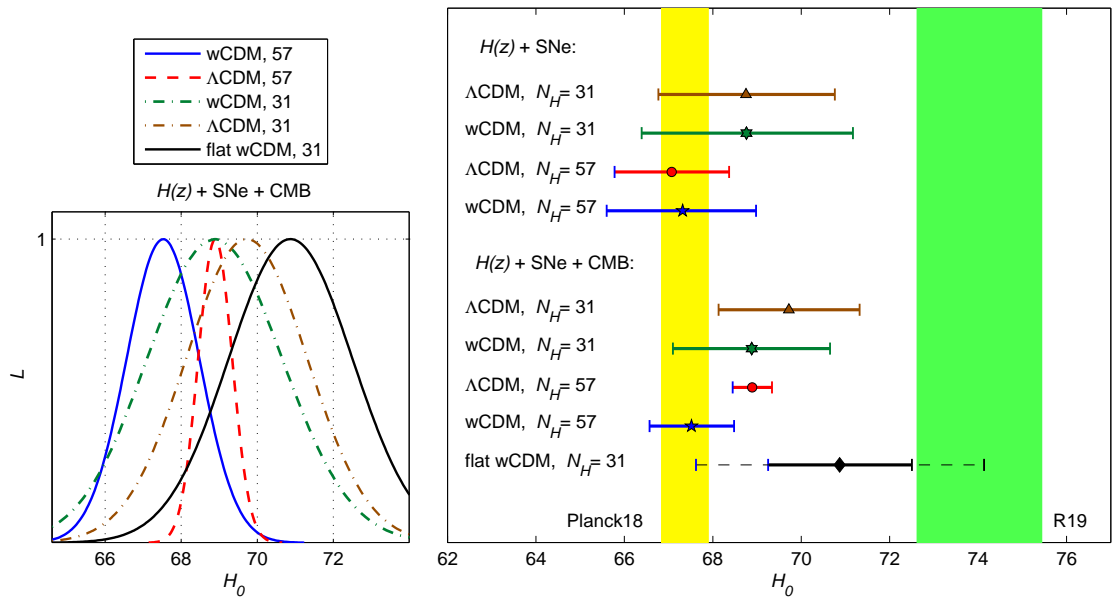}}
\caption{ Likelihoods for $\chi^2_{\mbox{\scriptsize tot}}$ ($H+{}$SNe Ia${}+{}$CMB)
with the correspondent whisker plot (with the previous case $\chi^2_{H+\mbox{\scriptsize
SN}}$) in comparison with  Planck18 and R19 $H_0$ estimates. The dashed line describes
the $2\sigma$ error band, solid thick lines correspond to  $1\sigma$ estimates.}
  \label{FWh2}
\end{figure}

Estimations of the Hubble constant $H_0$ (shown in the top-left panel of Fig.~\ref{F5})
and the correspondent whisker plot with $1\sigma$ error boxes
are presented in Fig.~\ref{FWh2}. Here we also compare
the results for $H+{}$SNe Ia${}+{}$CMB data with the previous estimates from the
function $\chi^2_{H+\mbox{\scriptsize SN}}$. One can note that the included CMB data
almost do not change the best fitted $H_0$ estimates (with the mentioned contraction
of their error boxes) for the  $w$CDM model, but the $H_0$ estimates, but the $H_0$ estimates
for the  $\Lambda$CDM model appear to be enlarged. However, this growth is too small for
describing the HST (R19) estimations, that could be a solution of the $H_0$ tension problem.

The most successful variant for solving this problem is to consider the flat case
($k=0$) of the $\Lambda$CDM or $w$CDM models. This variant is the particular case of
these models, if we just suppose $\Omega_k=0$ in our calculations.  The corresponding
result $H_0=70.87^{+1.63}_{-1.62}$ km\,/(s$\cdot$Mpc) for the flat $w$CDM model with
$N_H=31$ is shown in Fig.~\ref{FWh2} with black color. The $1\sigma$ band for this
variant is very close to R19 estimates, however only the correspondent $2\sigma$ band
(shown as the dashed line) reaches the  R19 range.

\section{Conclusion}

In this paper we considered two cosmological models $\Lambda$CDM and $w$CDM in
confrontation with different observational data: the Hubble parameter $H(z)$ estimations
(31 data points from cosmic chronometers and the extended sample with 57 data points),
the Pantheon sample  Type Ia supernovae data \cite{Scolnic17} and CMB data in the form
(\ref{chiCMB}), (\ref{CMBpriors}) \cite{HuangWW2015}. In this study we, in particular,
kept in mind a possibility to alleviate the Hubble constant tension between the Planck
\cite{Planck13,Planck15,Planck18}  and HST \cite{HST18,HST19} estimations of $H_0$.

We have shown that the $H_0$ tension can be easily explained (with simple ``switching''
from the $\Lambda$CDM to $w$CDM model), if we consider only the $H(z)$ data via the
$\chi^2_H$ function (\ref{chiH}) (see Figs.~\ref{F2}, \ref{FWh}). However, this approach
with the extremely poor set of observations is not acceptable, because it predicts
extraordinary values of model parameters in Table~\ref{Estim}.

The model predictions become reliable, when we include into consideration the SNe Ia
\cite{Scolnic17} and CMB data \cite{HuangWW2015}. The resulting best fitted values of
model parameters with $1\sigma$ errors for the $\chi^2$ functions
$\chi^2_{H+\mbox{\scriptsize SN}}$ (\ref{chiHSN}) and $\chi^2_{\mbox{\scriptsize tot}}=
\chi^2_{H+\mbox{\scriptsize SN}}+\chi^2_{\mbox{\scriptsize CMB}}$ (\ref{chitot}) are
presented in Table~\ref{Estim2}. The corresponding results for Hubble constant $H_0$ are
shown in Fig.~\ref{FWh2}, they essentially depend on chosen filters inside the models
(for example, if we fix  $w=-1$ or $\Omega_k=0$) or filters applied to observations.

 {\small
\begin{table}[th]
\caption{The best fitted values and $1\sigma$ estimates of model parameters for
$H(z)+{}$SN and CMB data}
\begin{center}
\begin{tabular}{||c|c|c|c|c|c|c|c||}
\hline Model & Data & $N_H$ &$\min\chi^2$
\rule{0pt}{1.4em}& $H_0$ & $\Omega_m^0$  & $\Omega_k$ & $w$  \\ \hline
$\Lambda$CDM& $H$+SN & $31$& $1072.76$ & $68.75^{+2.01}_{-1.98}$ & $0.322^{+0.066}_{-0.068}$ &  $-0.035^{+0.176}_{-0.167}$ &  $-1$ \rule{0pt}{1.4em}  \\
\hline
$w$CDM & $H$+SN & $31$ & $1072.76$ & $68.76^{+2.41}_{-2.37}$ &  $0.322^{+0.066}_{-0.069}$&  $-0.036^{+0.290}_{-0.258}$  &  {\footnotesize$-0.988^{+0.166}_{-0.32}$}\rule{0pt}{1.4em} \\
\hline
$\Lambda$CDM& $H$+SN & $57$&  $1088.76$ & $67.07^{+1.30}_{-1.29}$ & $0.242^{+0.027}_{-0.029}$ &  $0.170^{+0.096}_{-0.092}$ & $-1$ \rule{0pt}{1.4em}  \\
\hline
$w$CDM & $H$+SN & $57$ & $1088.70$ & $67.32^{+1.66}_{-1.72}$ &  $0.252^{+0.048}_{-0.061}$&  $0.108^{+0.30}_{-0.251}$  &  {\footnotesize$-0.954^{+0.124}_{-0.330}$}\rule{0pt}{1.4em} \\
\hline\hline
$\Lambda$CDM& {\footnotesize$ H$+SN+CMB}& $31$& $1074.29$ & $69.72^{+1.60}_{-1.59}$ &{\footnotesize$0.2829^{+0.0017}_{-0.0018}$} &  $0.0056^{+0.0017}_{-0.0017}$ & $-1$ \rule{0pt}{1.4em}  \\
\hline
$w$CDM & {\footnotesize$ H$+SN+CMB}& $31$& $1073.20$ & $68.88^{+1.77}_{-1.78}$ & {\footnotesize$0.2826^{+0.0017}_{-0.0018}$} & $0.009^{+0.004}_{-0.004}$ & {\footnotesize$-0.945^{+0.051}_{-0.053}$}\rule{0pt}{1.4em} \\
\hline
$\Lambda$CDM&{\footnotesize$ H$+SN+CMB}&$57$&  $1092.09$ & $68.89^{+0.45}_{-0.44}$ & {\footnotesize$0.2426^{+0.0017}_{-0.0017}$} &  $0.0055^{+0.0017}_{-0.0017}$ & $-1$ \rule{0pt}{1.4em}  \\
\hline
$w$CDM &{\footnotesize$ H$+SN+CMB}& $57$& $1089.44$ & $67.52^{+0.96}_{-0.95}$ &  {\footnotesize$0.2822^{+0.0017}_{-0.0018}$} & $0.011^{+0.004}_{-0.004}$ & {\footnotesize$\!-0.922^{+0.048}_{-0.048}\!$}\rule{0pt}{1.4em} \\
\hline
 \end{tabular}
\end{center}
 \label{Estim2}
\end{table}
}

If we concentrate on the $H_0$ tension problem, we may conclude that the most successful
scenario for its alleviation is the $w$CDM model with the maximal data set (for
$\chi^2_{\mbox{\scriptsize tot}}= \chi^2_{H+\mbox{\scriptsize SN}+\mbox{\scriptsize
CMB}}$): the best fitted value $H_0=67.52^{+0.96}_{-0.95}$ km\,s${}^{-1}$Mpc${}^{-1}$
for $N_H=57$ almost coincides with the Planck 18 estimate \cite{Planck18}; from the
other side, if we accept the flat variant of this model (fix $\Omega_k=0$) we obtain
$H_0=70.87^{+1.63}_{-1.62}$ km\,s${}^{-1}$Mpc${}^{-1}$ for $N_H=31$ that is very close
to the  HST estimation \cite{HST19} (the green band in Fig.~\ref{FWh2}), but it is not
large enough and lies outside the $1\sigma$ confidence level (only $2\sigma$ bands have
intersection).

One may conclude that the $w$CDM model has considerable achievements, but it is not
successful enough for conclusive solving the $H_0$ tension problem on the base of the
mentioned observational data. For this purpose we should investigate some its extensions
or other cosmological scenarios \cite{DiValentMLS2017}\,--\,\cite{DiValentMMVw2019}.

\end{document}